\documentclass[pra,superscriptaddress,twocolumn,amsmath,nofootinbib,longbibliography]{revtex4}
\usepackage{epsfig}
\usepackage{bm}

\newcommand{\vA}{\mathbf{A}}

\newcommand{\IP}{I_p}

\newcommand{\vks}{\vk_\mathrm{st}}
\newcommand{\ver}{\mathbf{r}}

\newcommand{\vT}{\mathbf{T}}
\newcommand{\vd}{\mathbf{d}}

\newcommand{\vk}{\mathbf{k}}

\newcommand{\ve}{\hat{\mathbf{e}}}

\newcommand{\vE}{\mathbf{E}}

\newcommand{\Wcm}{\;\mathrm{W/cm}^2}

\begin{document}
 \title{Control of the helicity of high-order harmonics generated by bicircular field}
\author{D. B. Milo\v{s}evi\'{c}}
 \affiliation{Faculty of Science, University of Sarajevo, Zmaja od Bosne 35, 71000 Sarajevo, Bosnia and Herzegovina}
 \affiliation{Academy of Sciences and Arts of Bosnia and Herzegovina, Bistrik 7, Sarajevo, Bosnia and Herzegovina}
 \affiliation{Max-Born-Institut, Max-Born-Strasse~2a, 12489 Berlin, Germany}
\date{\today}
\begin{abstract}
High-order harmonics generated by bicircular laser field have
helicities which alternate between $+1$ and $-1$. In order to
generate circularly polarized high-harmonic pulses, which are
important for applications, it is necessary to achieve asymmetry in
emission of harmonics having opposite helicities. We theoretically
investigated a wide range of bicircular field component intensities
and found areas where both the harmonic intensity is high and the
helicity asymmetry is large. We investigated the cases of
$\omega$--$2\omega$ and $\omega$--$3\omega$ bicircular fields and
atoms having the $s$ and $p$ ground states, exemplified by He and Ne
atoms, respectively. We have shown that for He atoms strong high
harmonics having positive helicity can be generated using
$\omega$--$3\omega$ bicircular field with a much stronger second
field component. For Ne atoms the helicity asymmetry can be large in
a wider range of the driving field component intensities and for
higher harmonic orders. For the stronger second field component the
harmonic intensity is higher and the helicity asymmetry parameter is
larger for higher harmonic orders. The results for Ne atoms are
illustrated with the parametric plots of elliptically polarized
attosecond high-harmonic field.
\end{abstract}

\maketitle

\section{Introduction}\label{sec:introduction}
High-order harmonic generation (HHG) is a
strong-laser-field-induced process in which the energy absorbed from
the laser field is emitted in the form of a high-energy photon. For
this process it is crucial that the electron, temporarily liberated
from an atom, moves away from it, turns around and returns to the
parent core to recombine with it emitting a high harmonic. This
process was discovered for a linearly polarized driving laser field
for which the emitted high harmonics are linearly polarized
\cite{McPherson:87,Ferray:JPB88}.

For specific applications it is important to generate circularly
polarized high harmonics. Such harmonics can be generated using the
so-called bicircular field which consists of two coplanar
counter-rotating circularly polarized fields having different
frequencies
\cite{Eichmann:PRARC95,Long:PRA95,Zuo:JNOPM95,Milosevic:PRA00}. That
the harmonics generated by bicircular field are circularly polarized
was confirmed in \cite{Fleischer:NatPhot14}. Since these harmonics
have helicities which alternate between $+1$ and $-1$, by combining
a group of such harmonics, instead of obtaining a circularly
polarized attosecond pulse train, one obtains a pulse with unusual
polarization properties (in \cite{Milosevic:PRARC00} a starlike
structure with three linearly polarized pulses rotated by
$120^\circ$ was predicted; this was confirmed experimentally in
\cite{Chen:ScienceAdvances16}). However, if the harmonics of
particular helicity are stronger, i.e., if we have helicity
asymmetry in a high-harmonic energy interval, then it is possible to
generate elliptic or even circular pulse train. Such pulses can then
be used to explore chirality-sensitive processes which, for example,
appear in magnetic materials \cite{Fan:PNAS15,Kfir:SciAdv17},
organic molecules \cite{Cireasa:NatPhys15,C6CP01293K} etc.
Fortunately, such a helicity asymmetry exists for HHG by inert gases
having the $p$ ground state
\cite{Milosevic:OL15,Mediauskas:PRL15,Milosevic:PRA15}.

The aim of the present work is to explore the influence of the
bicircular-laser-field-component intensities on the circularly
polarized high-order harmonics. Such an analysis shows how one can
control the ratio of the intensities of harmonics having opposite
helicities. It should be mentioned that, in addition to HHG (see
more recent
Refs.~\cite{Dorney:PRL17,Heslar:PRA17,Jimenez-Galan:PRA18,Frolov:PRL18}),
other strong-field processes in bicircular field have been explored
(for reviews see \cite{Odzak:JMO17,Bandrauk:JOptR17}). Examples are
above-threshold detachment \cite{Kramo:LPL07}, laser-assisted
electro-ion radiative recombination \cite{Odzak:PRA15}, high-order
above-threshold ionization
\cite{Mancuso:PRA15,Hasovic:OE16,Milosevic:PRA16,Mancuso:PRA16,Hoang:PRARC17,Milosevic:JPB18,Li:PRA18,Gazibegovic:OE18},
laser-assisted scattering \cite{Korajac:EPJD17}, nonsequential
double ionization \cite{Mancuso:PRL16,EckartDoerner:PRL16}, electron
vortices in photoionization \cite{NgokoDjiokap:PRL15,Pengel:PRL17},
spin-dependent effects \cite{Milosevic:PRARC16}, subcycle
interference effects \cite{Eckart:PRA18}, attoclock photoelectron
interferometry \cite{Han:PRL18}, high harmonics from relativistic
plasmas \cite{Chen:PRE18}, and optical chirality in nonlinear optics
\cite{Neufeld:PRL18}. The results of the present paper are relevant
also for these processes.

\section{Theory}\label{sec:theory}
We consider an $\omega$--$r\omega$ bicircular field with $r$
integer, the fundamental frequency $\omega=2\pi/T$, and the
component intensities $I_1=E_1^2$ and $I_r=E_r^2$, defined by (in
atomic units)
\begin{eqnarray}\label{Eq:1}
E_x(t)&=&\left[E_1\sin(\omega t)+E_r\sin(r\omega t)\right]/\sqrt{2},\nonumber\\
E_y(t)&=&\left[-E_1\cos(\omega t)+E_r\cos(r\omega
t)\right]/\sqrt{2}.
\end{eqnarray}
Our theory of HHG by bicircular field was presented in
\cite{Milosevic:PRA15}. The intensity of the $n$th harmonic is defined by
\begin{equation}\label{Eq:2}
I_n=\frac{(n\omega)^4}{2\pi c^3}\left|\vT_n\right|^2,\quad
\vT_n=\int_0^T \frac{dt}{T}\sum_m\vd_m(t)e^{in\omega t},
\end{equation}
where $\vd_m(t)$ is the time-dependent dipole and the magnetic
quantum number is $m=0$ for $s$ state or $m=\pm 1$ for $p$ state.
Using the dynamical symmetry of the bicircular field, one can derive
the following selection rule for the $n$th harmonic and its
ellipticity $\varepsilon_n$:
\begin{equation}
\varepsilon_n=\pm 1\;\;\mathrm{for}\;\; n=q(r+1)\pm 1.\;\; (q -
\mathrm{integer}) \label{Eq:3}
\end{equation}
Therefore, we can define the helicity asymmetry parameter by
\begin{equation}
A_{q(r+1)}=\frac{\left|T_{q(r+1)+1}\right|^2
-\left|T_{q(r+1)-1}\right|^2}{\left|T_{q(r+1)+1}\right|^2
+\left|T_{q(r+1)-1}\right|^2}.\label{Eq:12}
\end{equation}
The $T$-matrix element (\ref{Eq:2}) is calculated in the
strong-field approximation by integration over the recombination
time $t$, with
\begin{eqnarray}
\vd_m(t)&=& -i\left(\frac{2\pi}{i}\right)^{3/2}\int^\infty_0
\frac{d\tau}{\tau^{3/2}} \langle\psi_{ilm}|\ver|\vks +\vA(t)\rangle\nonumber\\
&&\times \langle \vks+\vA(t-\tau)|
\ver\cdot\vE(t-\tau)|\psi_{ilm}\rangle e^{i
S_\mathrm{st}}.\label{Eq:9}
\end{eqnarray}
In (\ref{Eq:9}) the integral is over the electron travel time $\tau$
of a product of the ionization and recombination matrix elements,
the electron wave-packet spreading factor $\tau^{-3/2}$, and a phase
factor with the action $S_\mathrm{st}\equiv -\IP\tau-\int_{t-\tau}^t
dt' \left[\vks+\vA(t')\right]^2/2$ in the exponent. Here
$\vE(t)=-d\vA(t)/dt$, $\vks\equiv -\int_{t-\tau}^t dt'\vA(t')/\tau$
is the stationary momentum, and $\IP$ the ionization potential. As
in \cite{Milosevic:PRA15,Milosevic:PRA18}, we model the atomic wave
function $\psi_{ilm}$ by a linear combination of the Slater-type
orbitals. All presented results, except those of Fig.~\ref{fig:2},
are focal-averaged over the laser intensity distribution. In all our
calculations we fix the fundamental wavelength to $\lambda=1300$~nm
(photon energy $\hbar\omega=0.9537$~eV). Results are calculated for
infinitely extended plane wave (CW) $T$-periodic laser field. It
should be mentioned that for a few-cycle laser pulse the harmonic
peaks are not well resolved and the results depend on the
carrier-envelope phase \cite{deBohan:PRL98}. In this case, for a
bicircular field, the results also depend on the relative phase
between the field components (i.e. on the time delay between the two
pulses) \cite{Frolov:PRL18}.

Focal-averaging over the laser intensity is done in the following
way. We suppose a Gaussian laser beam with the intensity
distribution in the focus given by
$I(r,z)=I_0\left(1+z^2/z_0^2\right)^{-1}\exp\{-r^2/[w_0^2(1+z^2/z_0^2)]\}$,
where $I_0$ is the peak intensity, $w_0$ is the minimum beam waist,
and $z_0=\pi w_0^2/\lambda$ is the Rayleigh range. The focal
averaged $n$th harmonic intensity is then obtained by integration
over all space:
$I_n(I_0)\propto\int_{-\infty}^{\infty}dz\int_0^{\infty}rdr
I_n(I(r,z))$. From this we get
\begin{equation}
I_n(I_0)\propto \int_0^{I_0}dI I_n(I)\sqrt{I_0-I}(2I+I_0)/I^{5/2}.
\end{equation}
This focal averaging is similar to that used for electrons in
ionization process. Instead of electrons we consider harmonic
photons. With this focal averaging we avoid oscillations in the
calculated helicity asymmetry parameter which appear for fixed
intensity. This focal averaging is an approximation since we do not
take into account macroscopic effects and the harmonic phase.

In the present paper we calculate harmonic intensity using
Eqs.~(\ref{Eq:2}) and (\ref{Eq:9}). The $T$-matrix element is
calculated by numerical integration over the times $t$ and $\tau$.
This two-dimensional integral over times can also be solved using
the saddle-point method. This approximation leads to the
quantum-orbit theory applied to HHG by bicircular field
\cite{Milosevic:PRA00,Milosevic:JMO18}. We shortly introduce some
elements of this theory which can be used to explain physical origin
and meaning of the obtained numerical results. In this theory the
$T$-matrix element is presented as $T_{0n}^j=\sum_sA_s^j e^{iS_s}$,
$j=x,y$, where the summation is over the solutions of the
saddle-point equations for the ionization time $t_{0s}$ and the
recombination time $t_s$, $S_s$ is the corresponding action, and the
amplitude $A_s^j$ is a product of the ionization, propagation, and
the recombination parts, known in analytical form. The times
$t_{0s}$ and $t_s$ are complex and can be used to calculate
two-dimensional electron trajectories and velocities projected into
the real plane. Analyzing partial contributions to the harmonic
intensity and the corresponding electron trajectories and velocities
one can better understand the HHG process in bicircular field as it
was done in Refs.~\cite{Milosevic:PRA00,Milosevic:JMO18}. For the
atoms with $p$ ground state one should also take into account
summation over the magnetic quantum number $m$ and the fact that the
corresponding matrix elements can be different for different values
of $m$. In Refs.~\cite{Milosevic:OL15,Milosevic:PRA15} a large
asymmetry in the $m=\pm 1$ contributions to the harmonic intensity
was used to explain the observed helicity asymmetry. An additional
explanation of this asymmetry is provided in
Ref.~\cite{Milosevic:PRA15} using a semiclassical model in which the
electron in the state $\psi_{ilm}$ is characterized with the
electron probability current density
$\mathbf{j}_m=m\left|\psi_{ilm}\right|^2\ve_\phi/r$. For HHG by
bicircular field, in order to be able to return to the parent ion to
recombine, the electron should have nonzero initial velocity $v_y$.
The larger is this velocity, the lower is the probability of
ionization. In the example presented in \cite{Milosevic:PRA15} it
was shown that for the ground state having $m=-1$ it is more
probable that the electron at the ionization time appears with a
larger initial velocity $\left|v_y\right|$ than in the $m=+1$ case,
which explains why the ionization probability and the harmonic
intensity are higher for the $m=-1$ case. In addition, for $m=\pm 1$
the recombination is such that the harmonic with the ellipticity
$\varepsilon_n=\pm 1$ are stronger.

\section{Numerical results for He atoms}\label{sec:numHe}
Let us first present results for He atom having the $s$ ground state
and the ionization potential $\IP=24.59$~eV. Since in this case the
magnetic quantum number is zero, we expect that the absolute value
of the helicity asymmetry parameter is small. Namely, in
\cite{Milosevic:OL15,Milosevic:PRA15} it was shown (and explained
using a semiclassical model) that the strong asymmetry between the
partial harmonic intensities for particular values of the magnetic
quantum number $m=\pm 1$ causes an asymmetry in emission of the
plateau harmonics having opposite helicities.

\begin{figure*}[t]
    \vspace{0.1cm} \centerline{
        \includegraphics[width=0.5\textwidth]{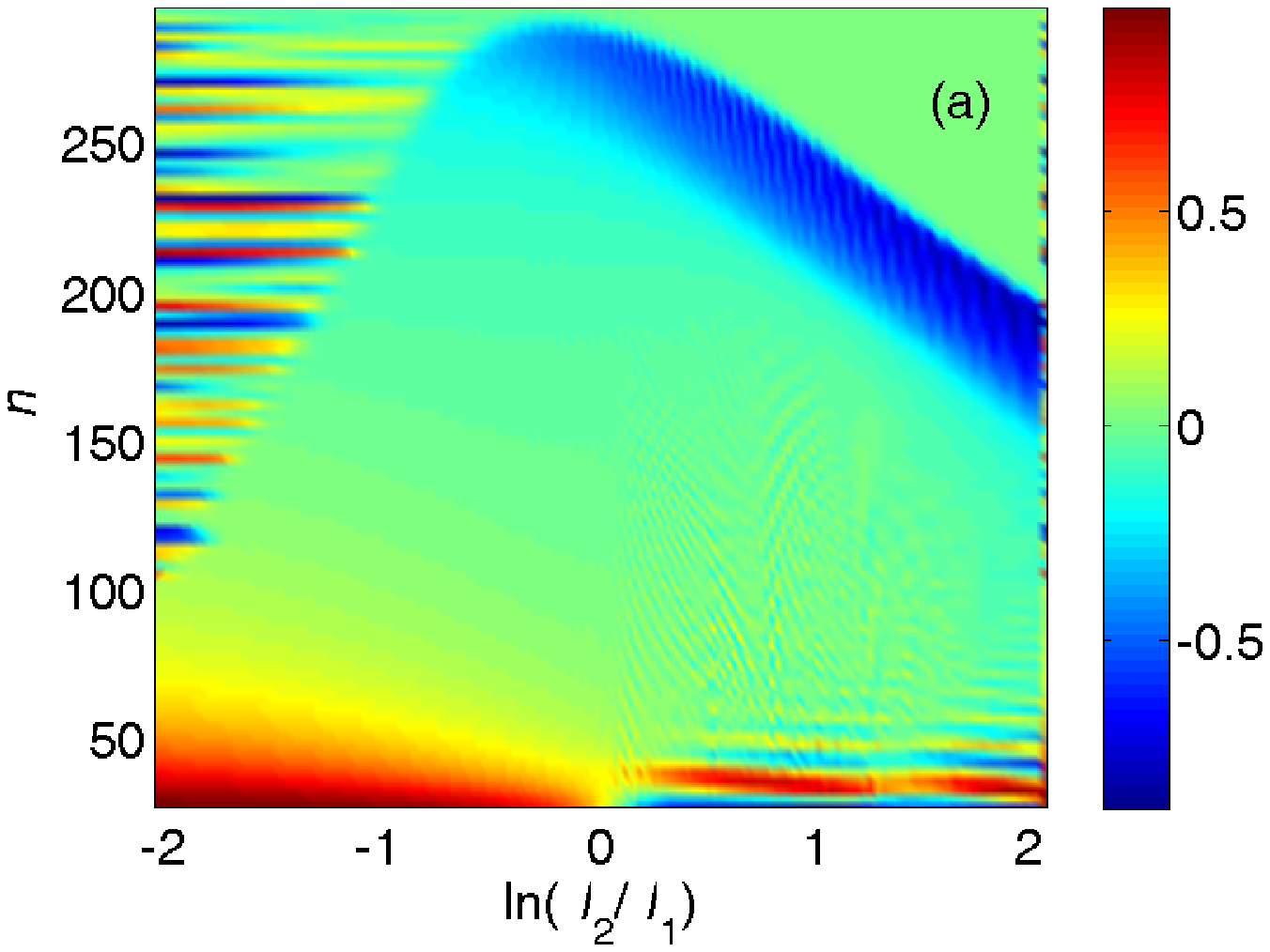}
        \quad
        \includegraphics[width=0.5\textwidth]{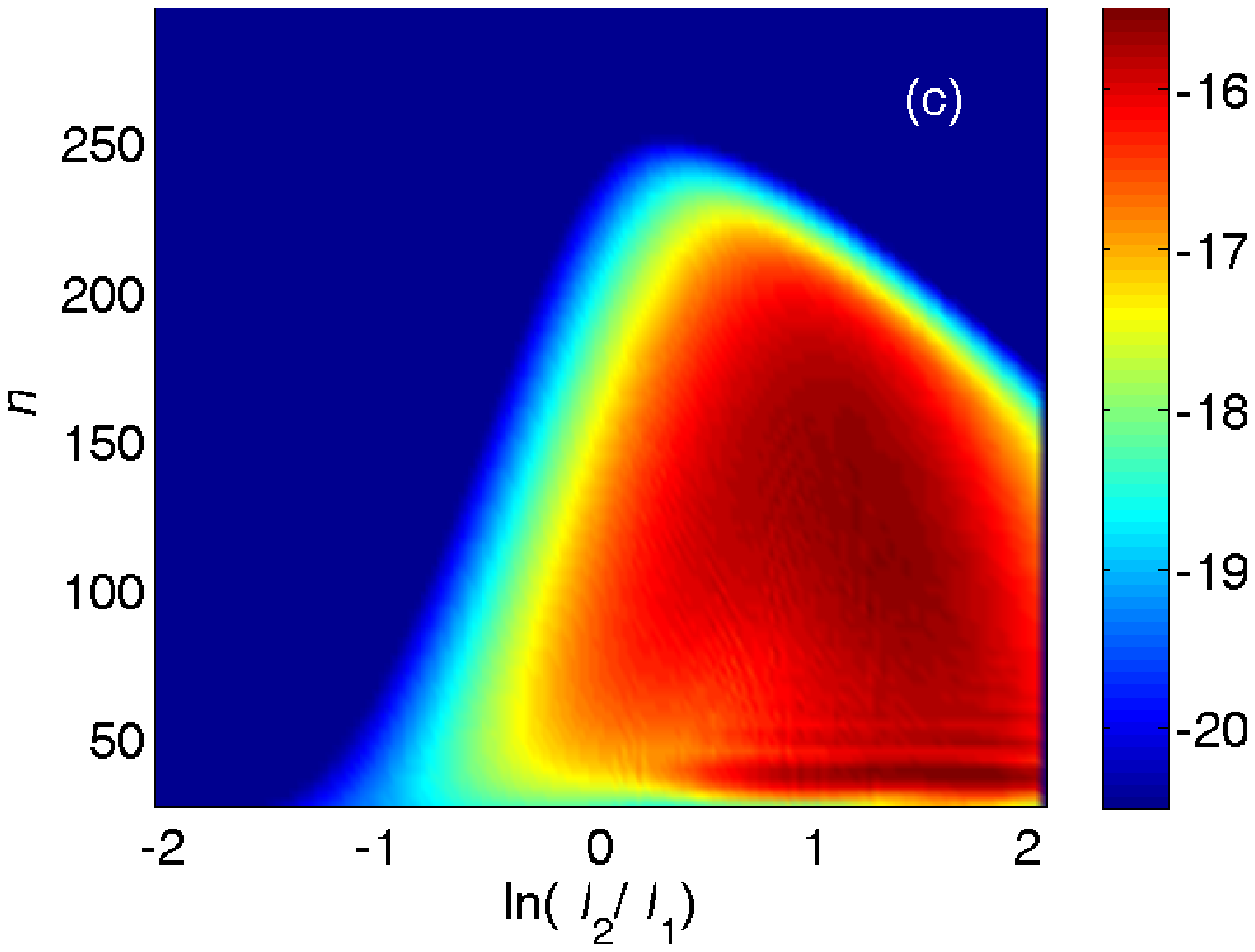}}
    \centerline{
        \includegraphics[width=0.5\textwidth]{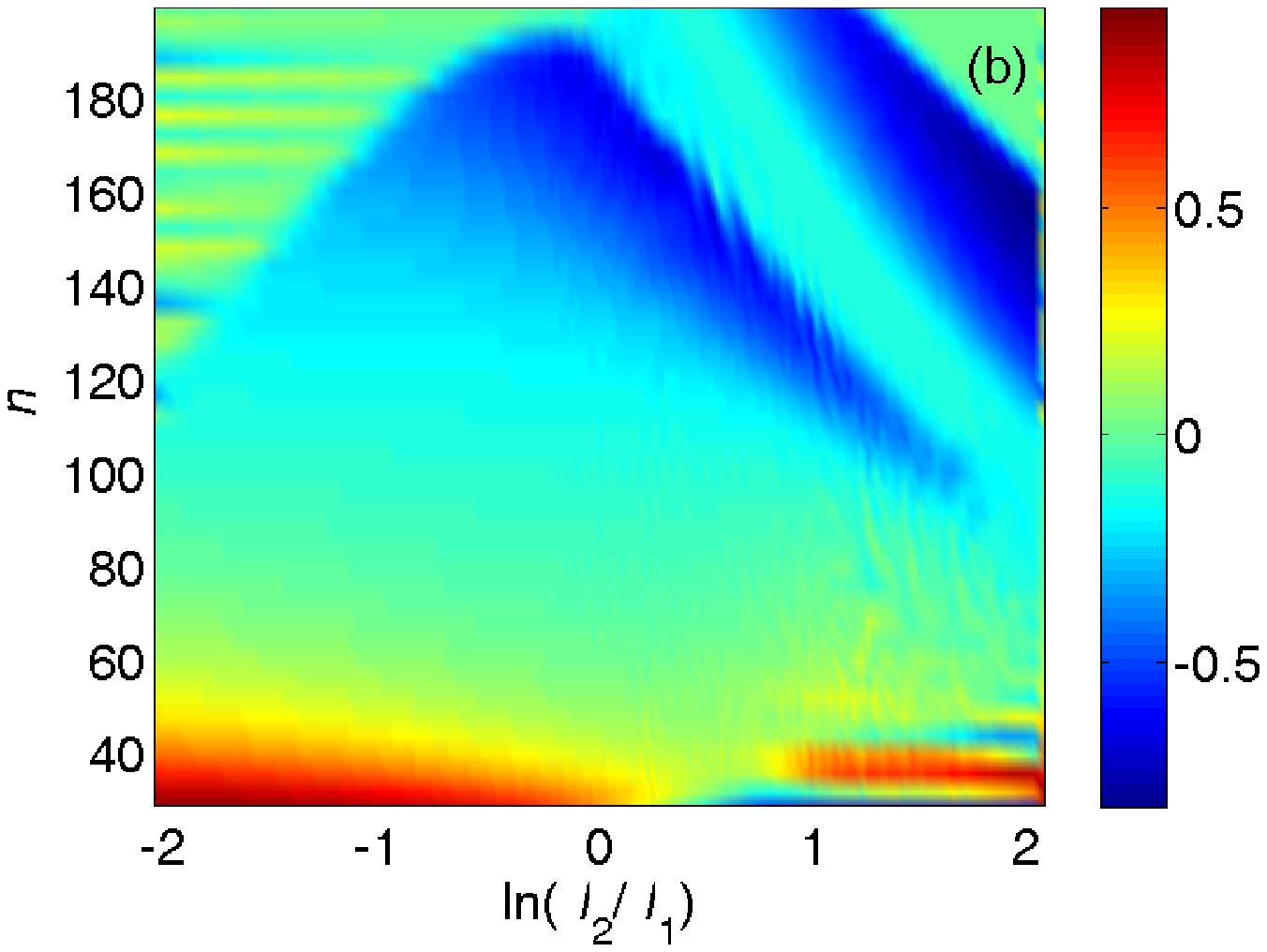}
        \quad
        \includegraphics[width=0.5\textwidth]{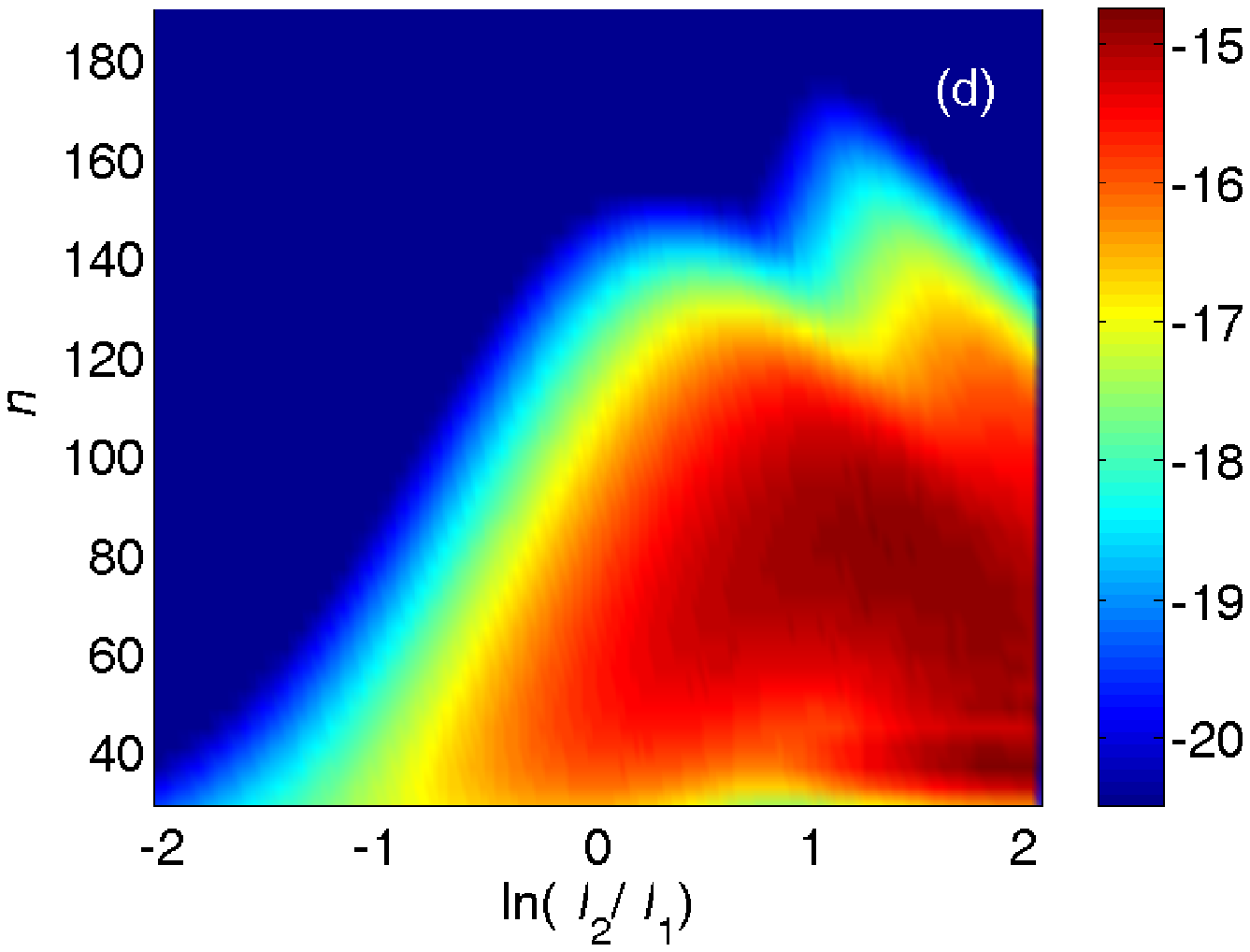}}
    \caption{Focal-averaged results for the helicity asymmetry parameter [left panels (a) and (b)]
and the logarithm of the harmonic intensity [right panels (c) and
(d)] for HHG by He atoms and $\omega$--$2\omega$ [upper panels (a)
and (c)] and $\omega$--$3\omega$ [lower panels (b) and (d)]
bicircular field with the fundamental wavelength 1300~nm. The
results are presented in false colors as a function of the natural
logarithm of the ratio of the component peak intensities,
$\mathrm{ln}(I_r/I_1)$, $r=2,3$, and of the harmonic order $n$. The
sum of the component peak intensities is fixed to $I_1+I_r=1.0\times
10^{15}\Wcm$.} \label{fig:1}
\end{figure*}
In Fig.~\ref{fig:1} we present focal-averaged results for the
helicity asymmetry parameter [panels (a) and (b)] and the logarithm
of the harmonic intensity [panels (c) and (d)] in false colors as
functions of the ratio of the bicircular field component peak
intensities and of the harmonic order $n$. The ratio $I_r/I_1$,
$r=2,3$, changes from $1/8$ to 8, while the presented natural
logarithm of this ratio, $\mathrm{ln}(I_r/I_1)$, changes from
$-2.0794$ to 2.0794. For the $\omega$--$2\omega$ case, presented in
the upper panels (a) and (c), the cutoff of the harmonic plateau is
the highest for $I_2\approx 2I_1$ and is above $n=200$. With the
decrease of the ratio $I_2/I_1$, the plateau length decreases as can
be seen in the panel (c). This manifests as a color structure in the
upper left part of the panel (a). Analogous is valid for $I_2>2I_1$
and the upper right part of the panel (a). The cutoff for the
$\omega$--$3\omega$ case, presented in the lower panels (b) and (d),
is above $n=160$ for $I_3> I_1$. It decreases with the decrease of
the ratio $I_3/I_1$. The mentioned change in the color structure in
the panel (b) nicely follows the shape of the cutoff region (the
helicity asymmetry in this region does not have a meaning since the
harmonics beyond the cutoff cannot be observed). The two blue
stripes in the upper right part of the panel (b) are connected with
the multiplateau structure which develops in the HHG spectrum for
the $\omega$--$3\omega$ case. This structure can be explained using
the quantum orbit theory and analyzing the corresponding partial
contributions to the harmonic emission rate \cite{Milosevic:JMO18}.
Namely, for $I_3>4I_1$, in addition to a plateau with the cutoff at
$n\approx 140$, a longer plateau appears with a cutoff above
$n=160$. The intensity of this longer plateau is lower and it is not
important for applications. The corresponding electron trajectories
have shape of a square while the electron velocity during the travel
time follows the shape of the vector potential
\cite{Milosevic:JMO18}. Partial harmonic intensities have their own
cutoffs which can be larger for longer orbits (this is contrary to
the linearly polarized field case for which the shortest orbits have
the highest cutoff).

What is the most important in Fig.~\ref{fig:1} is the large helicity
asymmetry for low harmonic orders. The red structures near $n=40$
appear both for the $\omega$--$2\omega$ and $\omega$--$3\omega$
cases. They are particularly noticeable for $I_r<I_1$ and diminish
for $I_r\approx I_1$. With the increase of the value of the ratio
$I_r/I_1$, the helicity asymmetry for $n<40$ appears again. It
appears earlier for the $\omega$--$2\omega$ case. For the
$\omega$--$3\omega$ case it becomes noticeable for $I_3>2I_1$.
Therefore, even using He atoms, having the $s$ ground state, it is
possible to generate highly chiral attosecond bursts by combining a
group of low harmonics. This was analyzed in detail in Sec.~4 of
Ref.~\cite{Milosevic:LP01} for $I_2=2I_1$. We have now shown that
this effect also exists for focal-averaged spectra and for other
intensity ratios. It is particularly pronounced for $I_1>I_r$.
Furthermore, a train of circularly polarized harmonic pulses can
also be generated using $\omega$--$3\omega$ bicircular field, as it
follows from Fig.~\ref{fig:1}(b). However, for applications it is
not enough that the helicity asymmetry parameter is large. It is
also necessary that the high-harmonic intensity is high.

\begin{figure}[t]
\includegraphics[width=\linewidth]{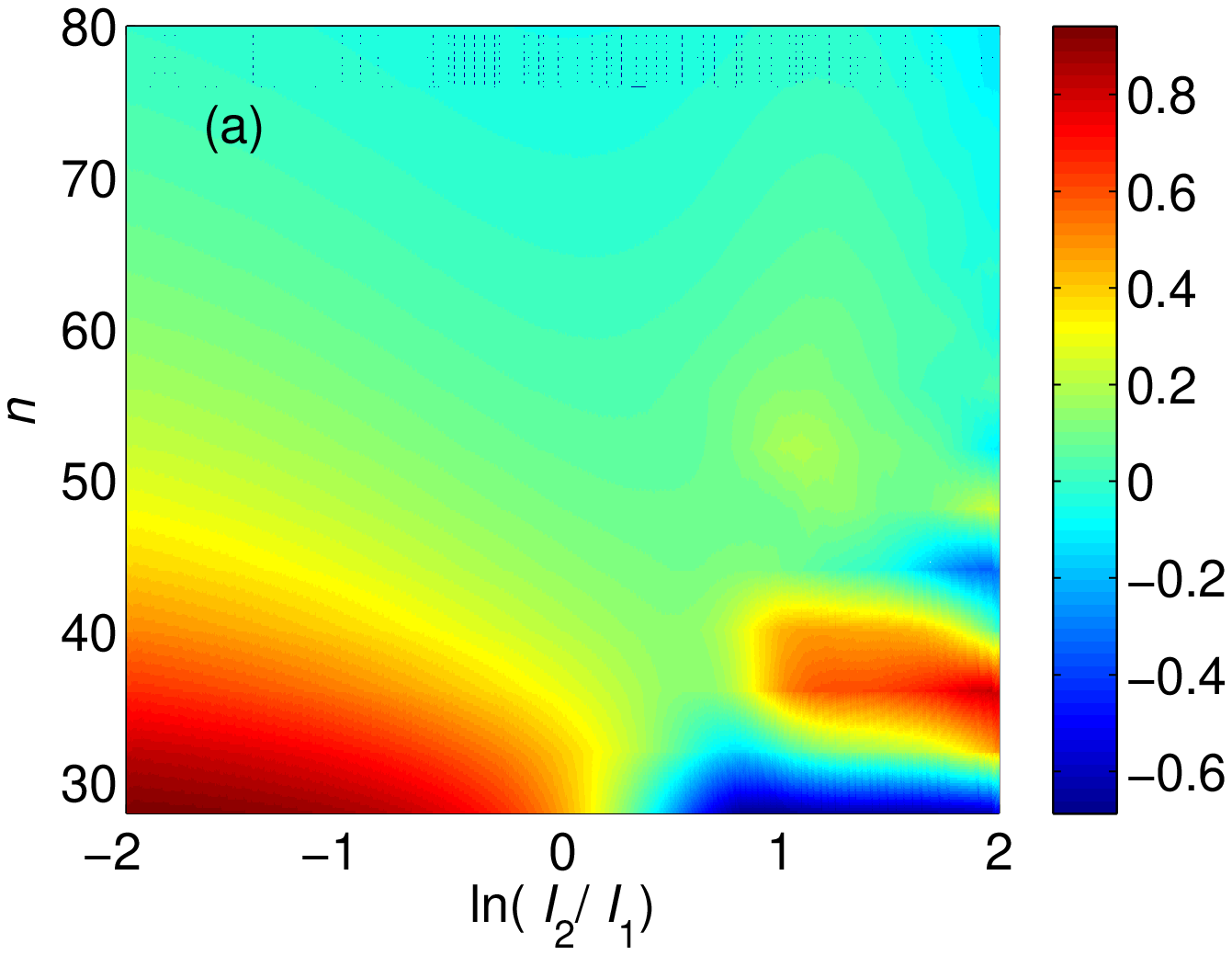}
\includegraphics[width=\linewidth]{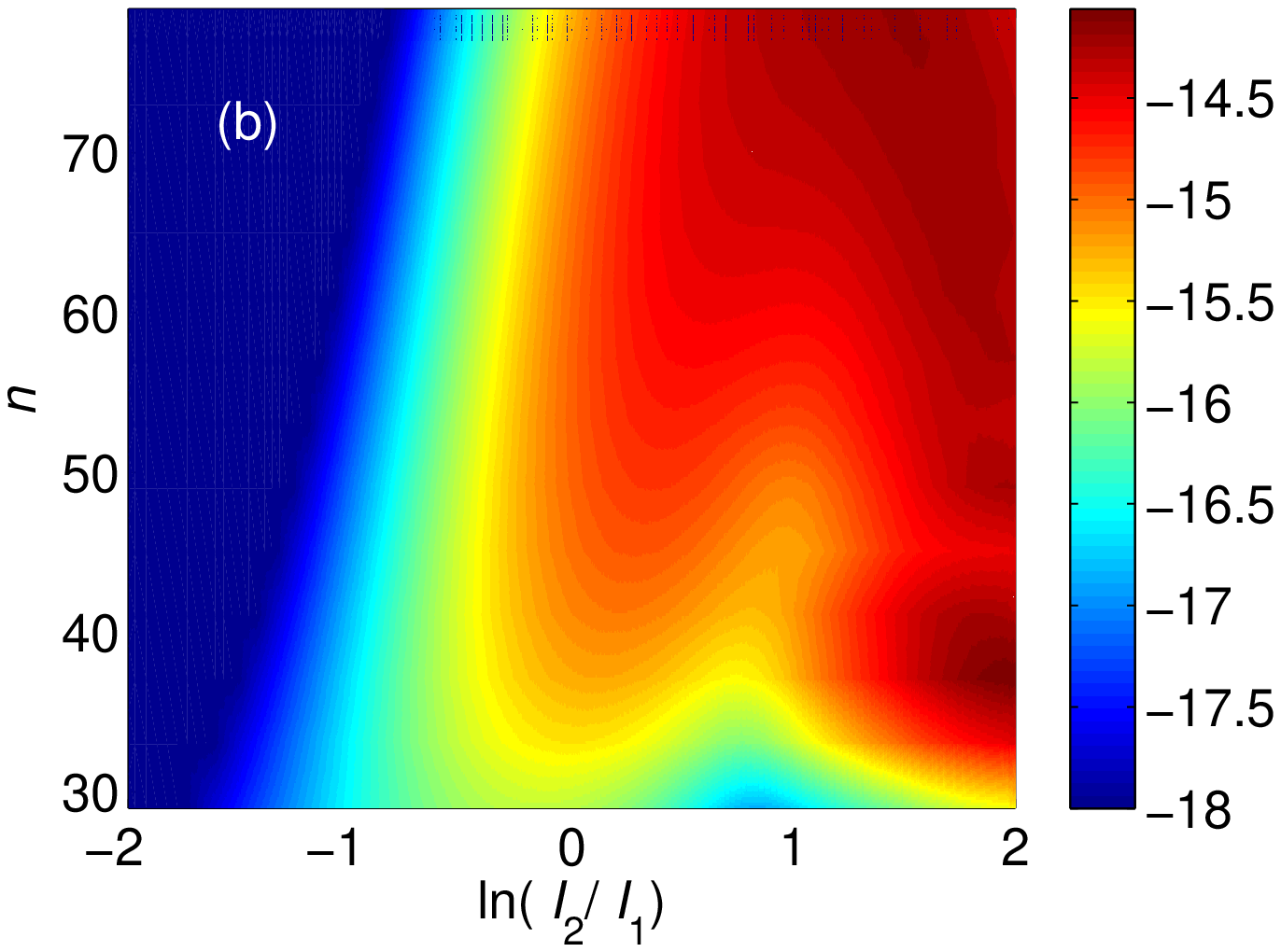}
\caption{Focal-averaged results for the helicity asymmetry parameter
[upper panel (a)] and the logarithm of the harmonic intensity [lower
panel (b)] for HHG by He atoms exposed to $\omega$--$3\omega$
bicircular field, presented in false colors as a function of the
natural logarithm of the ratio of the field component peak
intensities and harmonic order $n$. Other parameters are as in
Fig.~\ref{fig:1}.}\label{fig:1a}
\end{figure}
In  Fig.~\ref{fig:1a} we present the low-harmonic-order part of the
lower panels of Fig.~\ref{fig:1}, calculated with a higher
precision. We see that the intensity of harmonics for $I_3<0.5I_1$
is low so that, in spite of that the helicity asymmetry parameter is
large in this region, it is not important for practical application.
We also see that the most important is the case of $I_3>4I_1$ and
the region of the harmonic order near $n=40$. Therefore, strong high
harmonics of order $n>40$ having positive helicity can be generated
using $\omega$--$3\omega$ bicircular field with a much stronger
second field component ($I_3=8I_1$).

Using the formalism presented in the last paragraph of
Sec.~\ref{sec:theory} it can be shown \cite{Milosevic:JMO18} that
for $I_r\ge 2I_1$ the electron velocity at the ionization time is
small, while it is large for $I_r\le I_1$. The ionization
probability is higher for smaller velocity and this explains why the
harmonic intensity is higher for $I_r\ge 2I_1$ than for $I_r\le
I_1$. Furthermore, it was found in Ref.~\cite{Odzak:PRA15} that the
polarization of soft x-rays emitted in
bicircular-laser-field-assisted electron-ion radiative recombination
can be close to circular for low emitted x-ray photon energies and
for a wide range of incident electron angles. This favorizes the
emission of low harmonics having ellipticity $\varepsilon_n=+1$ and
large positive value of the helicity asymmetry parameter, exactly as
observed in Figs.~\ref{fig:1} and \ref{fig:1a}. An alternative
explanation why the helicity asymmetry parameter can be different
from zero for the $s$ ground state was recently presented in
\cite{Jimenez-Galan:PRA18}. It is based on the so-called propensity
rules and an analysis of the recombination matrix element. In this
paper a large helicity asymmetry parameter was noticed for the
$\omega$--$2\omega$ bicircular field and for higher intensity of the
first component. In our paper we found this effect also for the
$\omega$--$3\omega$ bicircular field. Furthermore, this effect
appears also for the stronger second field component. This case is
more important since the corresponding harmonic intensity is higher.

\section{Numerical results for Ne atoms}\label{sec:numNe}
\begin{figure}[t]
\includegraphics[width=\linewidth]{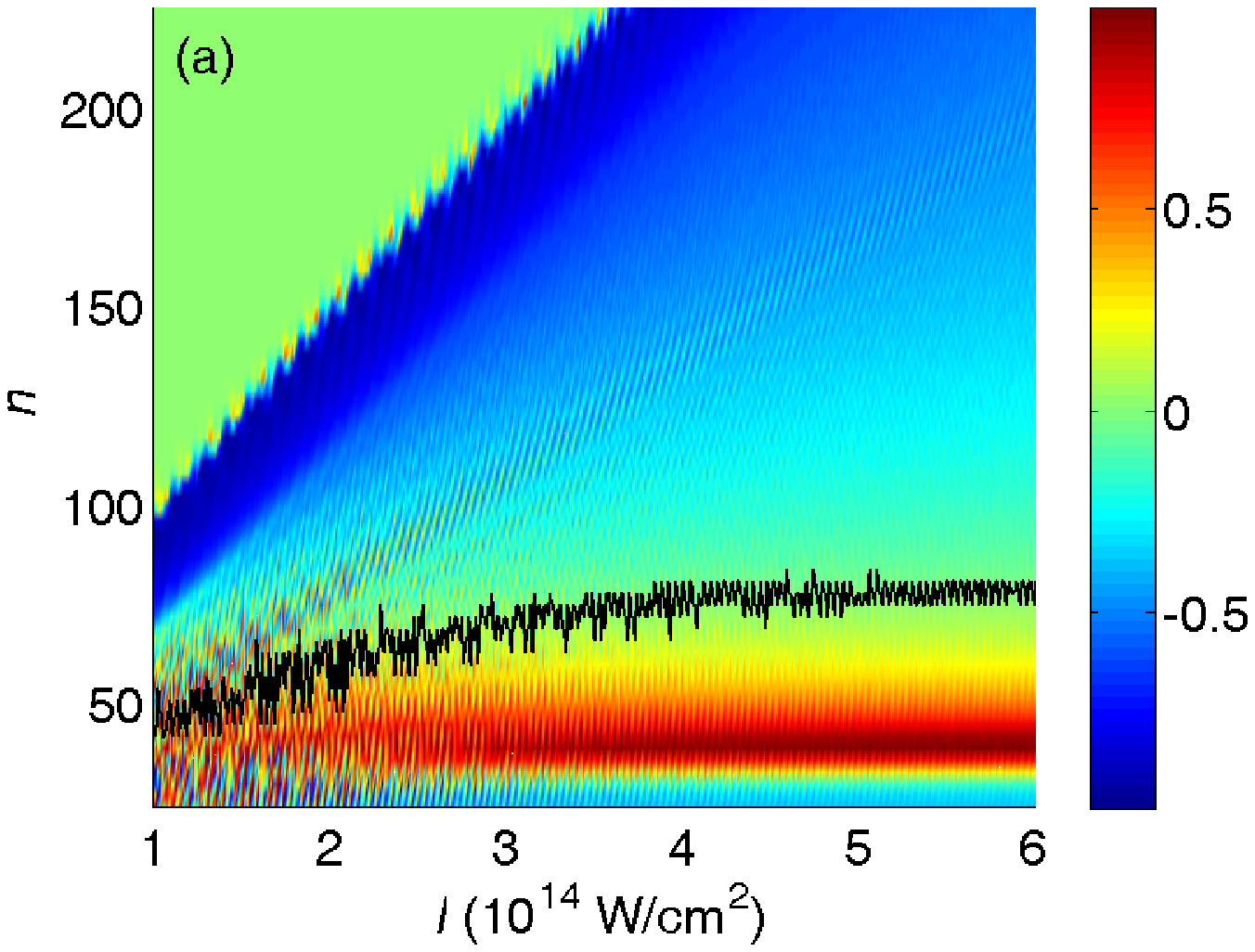}
\includegraphics[width=\linewidth]{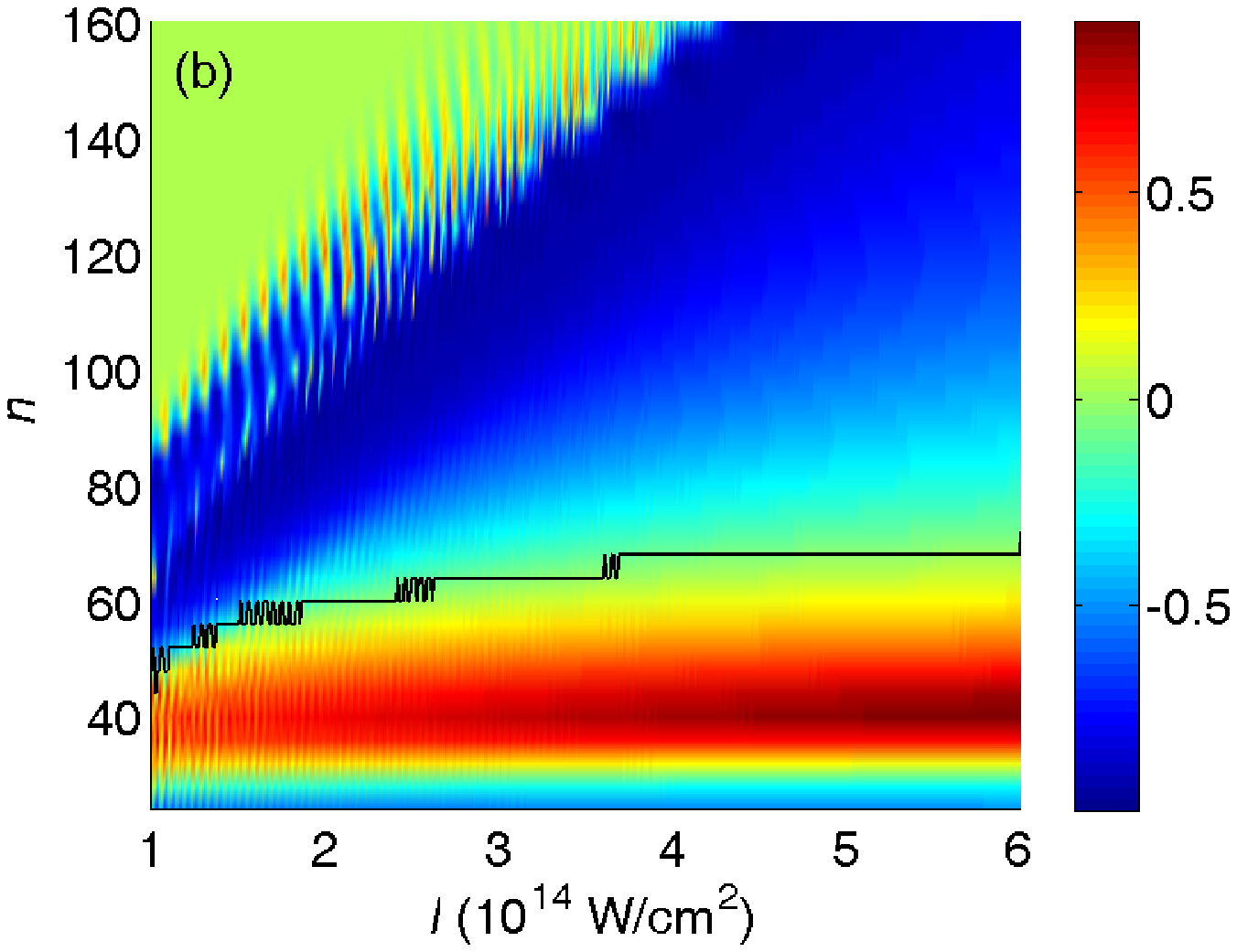}
\caption{Helicity asymmetry parameter presented in false colors as a
function of the laser-field intensity $I=I_1=I_r$, $r=2,3$, and
harmonic order $n$ for HHG by Ne atoms exposed to
$\omega$--$2\omega$ [upper panel (a)] and $\omega$--$3\omega$ [lower
panel (b)] bicircular field with the fundamental wavelength 1300~nm.
Black solid curve connects the point in which the asymmetry
parameter changes the sign. }\label{fig:2}
\end{figure}
\begin{figure*}[t]
    \vspace{0.1cm} \centerline{
        \includegraphics[width=0.5\textwidth]{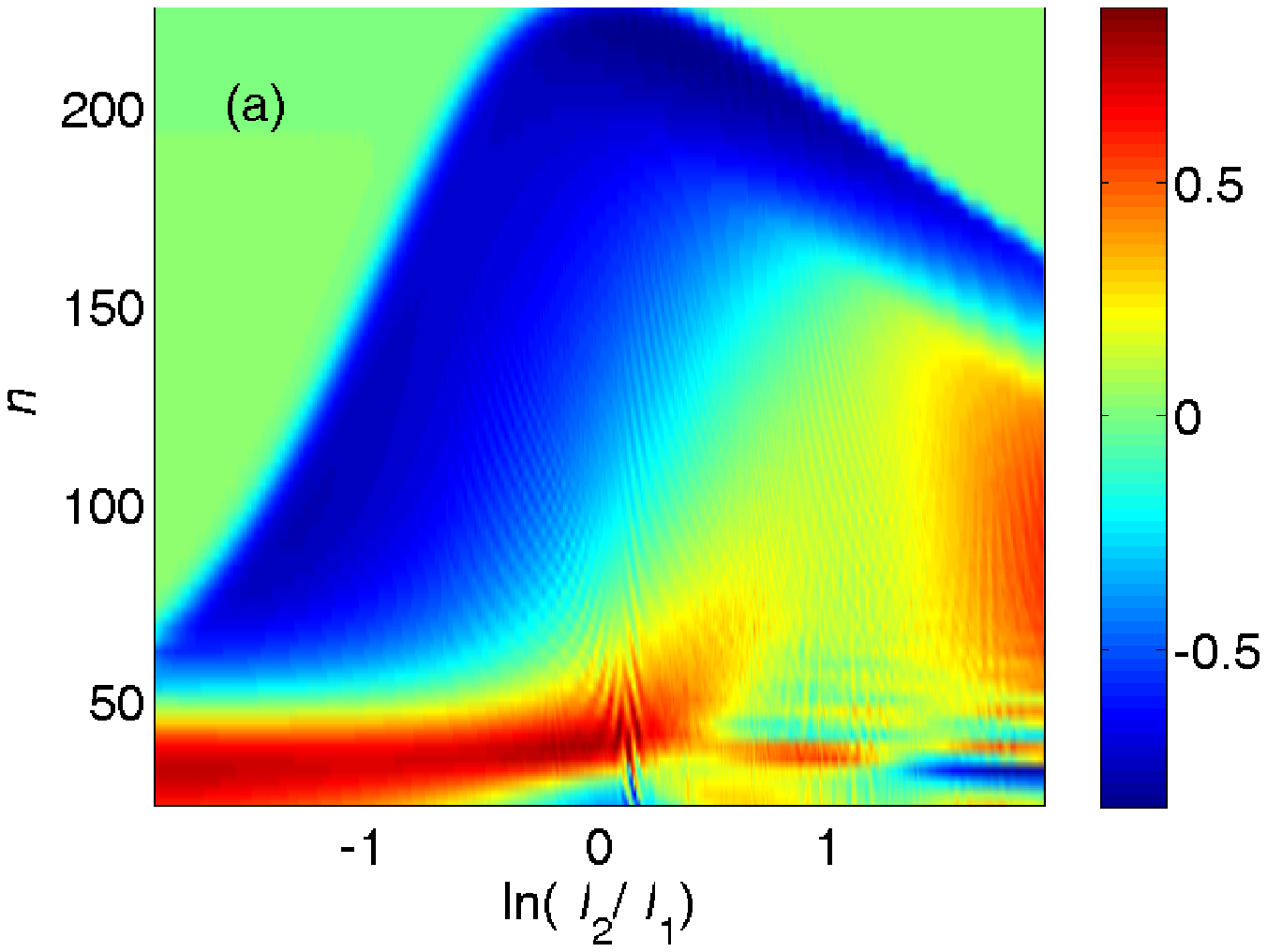}
        \quad
        \includegraphics[width=0.5\textwidth]{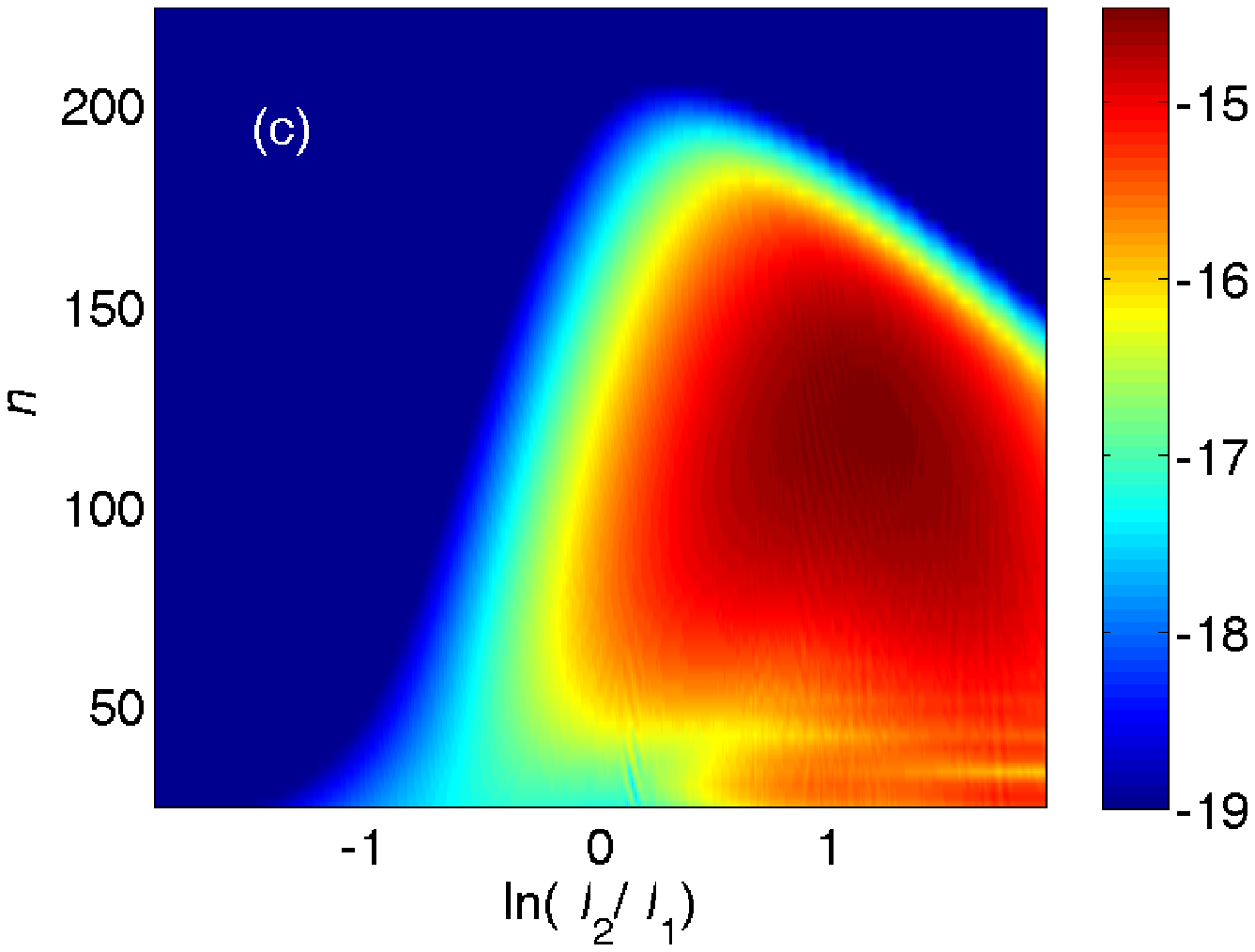}}
    \centerline{
        \includegraphics[width=0.5\textwidth]{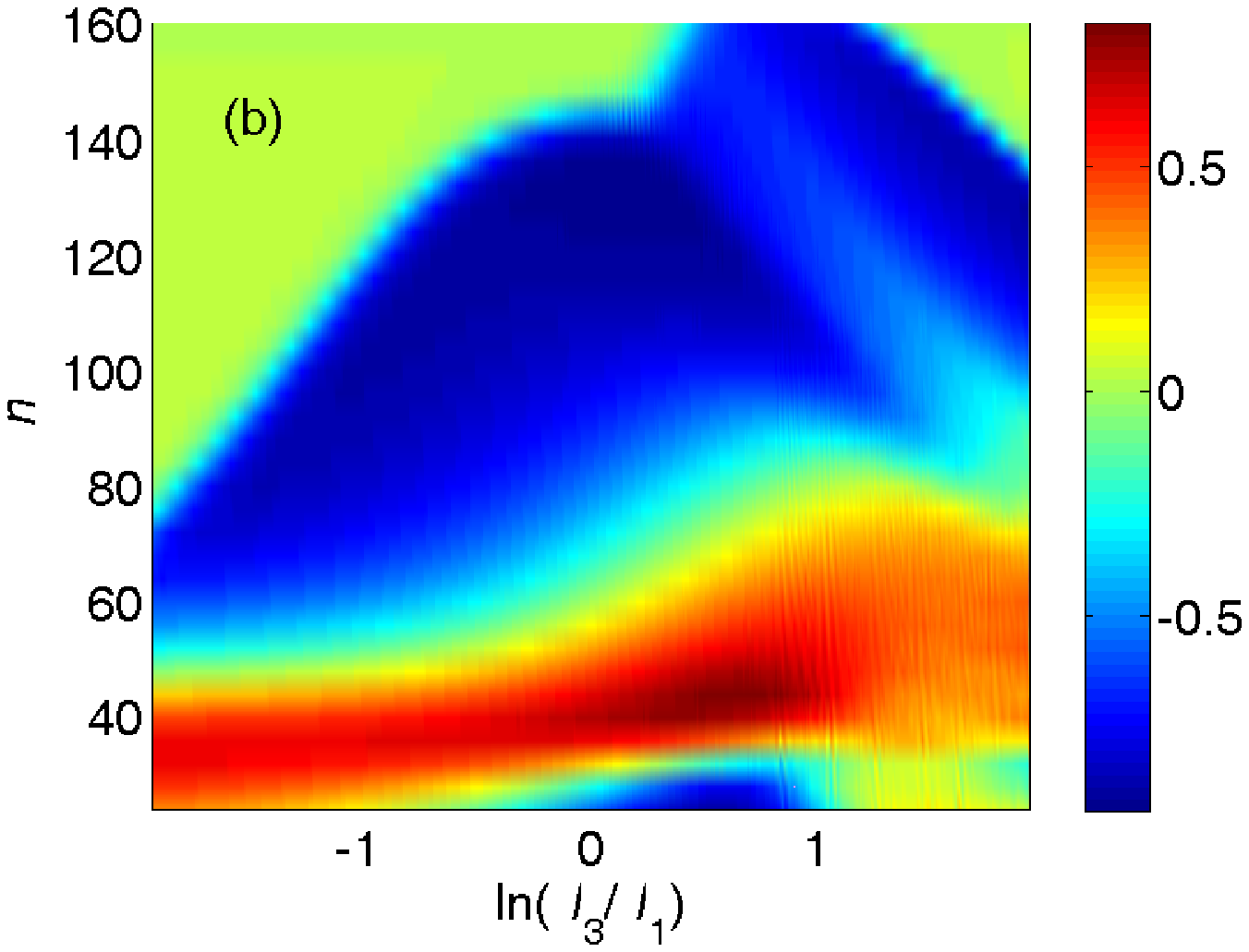}
        \quad
        \includegraphics[width=0.5\textwidth]{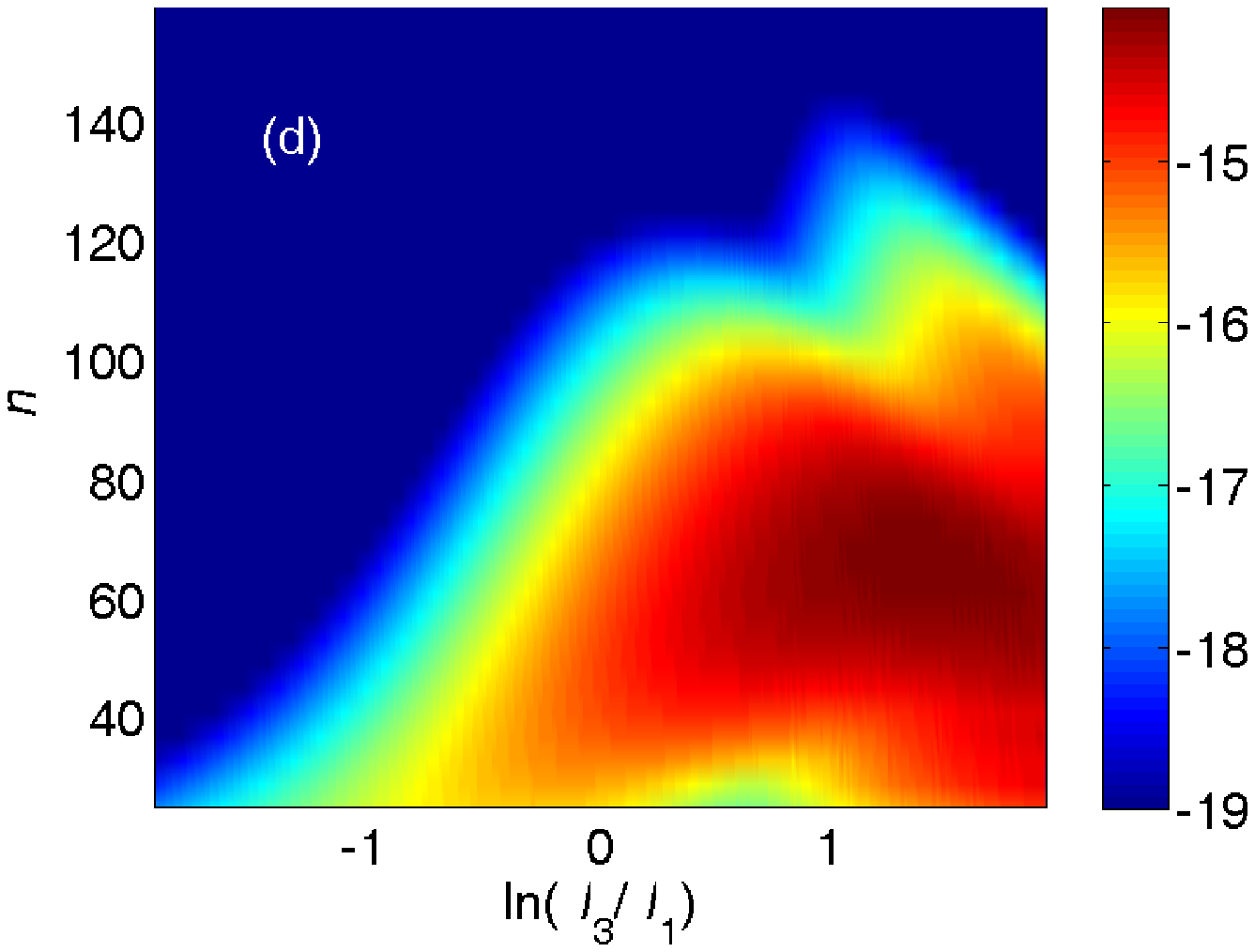}}
    \caption{Focal-averaged results for the
    helicity asymmetry parameter [left panels (a) and (b)]
and the logarithm of the harmonic intensity [right panels (c) and
(d)] for HHG by Ne atoms and $\omega$--$2\omega$ [upper panels (a)
and (c)] and $\omega$--$3\omega$ [lower panels (b) and (d)]
bicircular field with the fundamental wavelength 1300~nm. The
results are presented in false colors as a function of the natural
logarithm of the ratio of the component peak intensities,
$\mathrm{ln}(I_r/I_1)$, $r=2,3$, and of the harmonic order $n$. The
sum of the component peak intensities is fixed to $I_1+I_r=8\times
10^{14}\Wcm$.} \label{fig:3}
\end{figure*}
For Ne atoms with the $p$ ground state and the ionization potential
$\IP=21.56$~eV the situation is different. In Fig.~\ref{fig:2} we
present the helicity asymmetry parameter for equal component
intensities and without focal averaging. The cutoff region for the
$\omega$--$2\omega$ case [upper panel (a)] extends from $n=80$ for
$I_1=I_2=I=1\times 10^{14}\Wcm$ to above $n=200$ for $I>4\times
10^{14}\Wcm$. For the $\omega$--$3\omega$ case [lower panel (b)] the
cutoff is above $n=130$ for $I_1=I_3=I>4\times 10^{14}\Wcm$. For low
harmonic orders the harmonics with the positive helicities are
dominant, while for the plateau and cutoff harmonics we have the
opposite situation. For low laser-field intensities the helicity
asymmetry changes the sign near $n=50$. The corresponding harmonic
order $n$ increases with the increase of the laser intensity and for
the highest presented intensity it is $n=78$ for the
$\omega$--$2\omega$ case and $n=68$ for the $\omega$--$3\omega$
case. In all cases there is a wide region of the laser intensities
and harmonic orders for which the asymmetry parameter is large, so
that the HHG from Ne atoms can be used to explore chirality
sensitive processes.

Explanation why the helicity asymmetry parameter is large for atoms
having $p$ ground state is given in
Refs.~\cite{Milosevic:OL15,Milosevic:PRA15}. The reason is the
asymmetry in the $m=+1$ and $m=-1$ contributions to the
quantum-mechanical time-dependent dipole $\mathbf{d}_m(t)$. The
asymmetry in recombination with emission of harmonics having
ellipticity $\varepsilon_n=+1$ and $\varepsilon_n=-1$ is also
important. In terms of quantum orbits, semiclassical explanation is
connected with the electric ring current for atomic orbitals having
$m\ne 0$ whose sign is determined by the sign of $m$. As it is
explained at the end of Sec.~\ref{sec:theory}, it is more probable
that the electron appears with large velocity $|v_y(t_0)|$ for
$m=-1$ than for $m=+1$, which leads to higher ionization probability
and higher partial harmonic intensity. This is the reason for the
asymmetry of the $m=\pm 1$ contributions to the harmonic intensity
and for the large helicity asymmetry parameter for Ne. For lower
harmonic orders, as in the case of He atoms, the recombination with
emission of harmonics having the ellipticity $\varepsilon_n=+1$ is
more probable.

In Fig.~\ref{fig:3} we explore how the helicity asymmetry parameter
and the logarithm of the harmonic intensity for Ne atoms change with
the change of the ratio of the bicircular field component peak
intensities. The results are focal averaged and the sum of the
component peak intensities is fixed to $I_1+I_r=8\times
10^{14}\Wcm$, $r=2,3$. Ratio of the peak intensities $I_r/I_1$
changes from $1/7$ to 7, i.e., the natural logarithm
$\mathrm{ln}(I_r/I_1)$ changes from $-1.94591$ to $1.94591$. The
cutoffs of the HHG spectra are clearly visible in panels (c) and
(d). For low values of $I_r/I_1$ the HHG plateau does not develop at
all, i.e., the harmonic intensity falls quickly with the increase of
the harmonic order. With the increase of $I_r/I_1$ the plateau
starts to develop and for $r=2$ it is the longest slightly above
$I_2=I_1$. It is interesting  that for the $\omega$--$3\omega$ case
the plateau is very long even for $I_3=7I_1$. For application it is
important that in the region $I_r>2I_1$ very strong harmonics of the
order $100<n<150$ ($50<n<90$) for the $\omega$--$2\omega$
($\omega$--$3\omega$) bicircular field can be generated. From the
left panels (a) and (b) of Fig.~\ref{fig:3} we see that for
$I_r<I_1$ the harmonics having positive helicities are dominant for
low harmonic orders ($n<50$). This is similar to the case of He
atoms presented in Fig.~\ref{fig:1}. For Ne atoms and $n>50$, the
harmonics having negative helicity become dominant. With the
increase of the ratio $I_r/I_1$ the position of the zero helicity
asymmetry moves to higher harmonic orders. For the
$\omega$--$3\omega$ case particularly interesting is the region of
the harmonic order $40<n<50$ near $I_3=2I_1$, where
$0<\mathrm{ln}(I_3/I_1)<1$. In this region the asymmetry parameter
is close to 1 and the corresponding harmonic intensity is high. With
a further increase of the ratio $I_r/I_1$, both for the
$\omega$--$2\omega$ and $\omega$--$3\omega$ cases, the helicity
asymmetry parameter is large for a wide range of harmonic orders.
Furthermore, for the $\omega$--$2\omega$ case the harmonic intensity
is high in the region determined by $I_2>3I_1$ and $50<n<150$, while
for the $\omega$--$3\omega$ case the corresponding region is
determined by $I_3>I_1$ and $30<n<80$.

\begin{figure}[t]
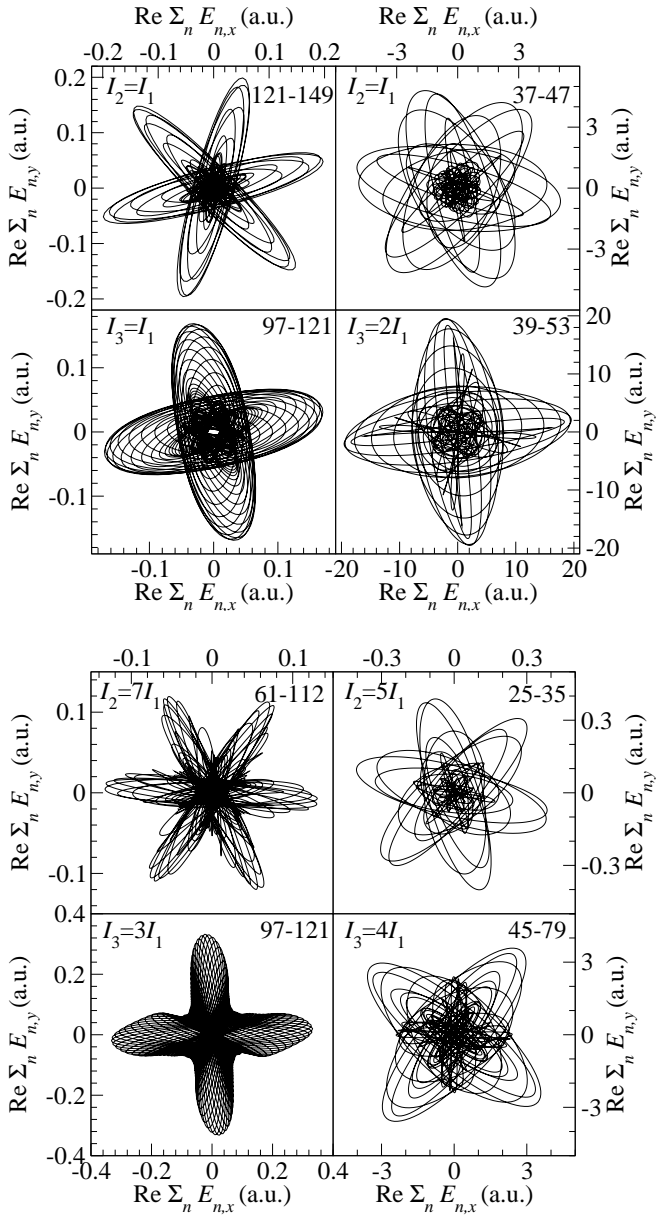

\includegraphics[width=\linewidth]{HarmonicField1_1.eps}\vspace{0.4cm}
\includegraphics[width=\linewidth]{HarmonicField2_1.eps}
\caption{Harmonic field of a group of harmonics from $n_1$ to $n_2$,
with the values of $n_1$ and $n_2$ denoted in the upper right corner
of each subpanel. High harmonics are generated by Ne atoms exposed
to $\omega$--$2\omega$ or $\omega$--$3\omega$ bicircular field. The
fundamental laser wavelength is 1300~nm and the sum of the component
peak intensities is fixed to $I_1+I_r=8\times 10^{14}\Wcm$, $r=2,3$.
Results are presented for different ratios of the intensities as
denoted in the upper left corner of each subpanel.}\label{fig:4}
\end{figure}

In Refs.~\cite{Milosevic:OL15,Milosevic:PRA15} we have shown that,
using a group of high harmonics generated by Ne atoms exposed to
bicircular field, it is possible to generate elliptically polarized
attosecond pulse trains. For this purpose we introduced the complex
time-dependent $n$th harmonic electric field vector $\vE_n(t)=n^2
\vT_n\exp(-i n\omega t)$, where $t$ is the harmonic emission time,
and considered the field formed by a group
of subsequent harmonics from $n=n_1$ to $n=n_2$. Let us now analyze
such high harmonic field for different values of the bicircular
field component frequencies and intensities and using focal
averaging. In Fig.~\ref{fig:4} we present, for one driving-field
optical cycle $T$, the parametric plot of the electric field vector
for various groups of harmonics. For the zero helicity asymmetry
parameter we would obtain a star-like structure
\cite{Milosevic:PRARC00} which consists of three (for
$\omega$--$2\omega$ field) or four (for $\omega$--$3\omega$ field)
approximately linearly polarized pulses. This threefold (fourfold)
symmetry reflects the corresponding symmetry of the driving field.
Since in our case the helicity asymmetry parameter is different from
zero we should obtain elliptically polarized attosecond pulse
trains. This is clearly visible in Fig.~\ref{fig:4}. The examples
presented comprise different groups of harmonics and different
ratios of the driving-field component intensities. Both the
$\omega$--$2\omega$ (intensities from $I_2=I_1$ to $I_2=7I_1$) and
$\omega$--$3\omega$ (intensities from $I_3=I_1$ to $I_3=4I_1$) cases
are included. Since the results are presented in the same units for
all panels, one can estimate the strengths of the corresponding
harmonic fields for all presented examples. The harmonic field
intensity is the highest for the case $I_3=2I_1$ and the group of 10
harmonics from $n_1=39$ to $n_2=53$
($n=40,41,43,44,46,47,49,50,52,53$).

\section{Conclusions}\label{sec:con}
In conclusion, for $\omega$--$2\omega$ and $\omega$--$3\omega$
bicircular fields we have explored a wide range of ratios of the
bicircular field component peak intensities $I_2/I_1$ and $I_3/I_1$
(from $1/8$ to 8 for He and from $1/7$ to 7 for Ne), with the goal
to find the laser intensity and harmonic order regions in which the
harmonic intensity is high and, at the same time, the high-harmonic
helicity asymmetry parameter is large. We hope that the presented
results will help the experimentalists in designing their
experiments for exploration of the chirality sensitive processes.

In particular, for He atoms, having the $s$ ground state, we have
shown that strong high-order harmonics of positive helicity can be
generated with the $\omega$--$3\omega$ bicircular field having a
much stronger second field component. For Ne atoms, having the $p$
ground state, the helicity asymmetry parameter can be large for
higher harmonic orders and in a wider range of the driving field
component intensities. We confirmed that the corresponding
high-harmonic pulses are elliptically polarized by presenting
parametric plots of the high-harmonic field. Physical explanation of
the obtained results is based on the quantum-orbit formalism.

\bibliography{BicirHHGSPM2018bib}

\end{document}